# Hybrid Quantum Deep Learning Model for Emotion Detection using raw EEG Signal Analysis

Ali Asgar Chandanwala, Srutakirti Bhowmik, Parna Chaudhury, Sheena Christabel Pravin*

School of Electronics Engineering,

Vellore Institute of Technology, Chennai.

*sheenachristabel.p@vit.ac.in

***Abstract:*** *Applications in behavioural research, human-computer interaction, and mental health depend on the ability to recognize emotions. In order to improve the accuracy of emotion recognition using electroencephalography (EEG) data, this work presents a hybrid quantum deep learning technique. Conventional EEG-based emotion recognition techniques are limited by noise and high-dimensional data complexity, which make feature extraction difficult. To tackle these issues, our method combines traditional deep learning classification with quantum-enhanced feature extraction. To identify important brain wave patterns, Bandpass filtering and Welch method are used as preprocessing techniques on EEG data. Intricate inter-band interactions that are essential for determining emotional states are captured by mapping frequency band power attributes (delta, theta, alpha, and beta) to quantum representations. Entanglement and rotation gates are used in a hybrid quantum circuit to maximize the model's sensitivity to EEG patterns associated with different emotions. Promising results from evaluation on a test dataset indicate the model's potential for accurate emotion recognition. The model will be extended for real-time applications and multi-class categorization in future study, which could improve EEG-based mental health screening instruments. This method offers a promising tool for applications in adaptive human-computer systems and mental health monitoring by showcasing the possibilities of fusing traditional deep learning with quantum processing for reliable, scalable emotion recognition.*

## 1. INTRODUCTION

Our emotional, psychological, and social well-being are all part of our mental health, which is a subset of behavioral health [1]. A condition of well-being known as mental health allows us to manage life's stressors, reach our full potential, study and work effectively, and give back to our community [2]. Mental illnesses, psychosocial impairments, and other mental states linked to severe suffering, functional impairment, or self-harm risk are all considered mental health issues. Although this is not always or always the case, people with mental health disorders are more likely to have poorer levels of mental well-being [3]. As a result, it is crucial to treat and care for those who are experiencing mental health issues.

Finding therapy for a mental health condition begins with a diagnosis. In the absence of a precise diagnosis, the individual can be given ineffective therapy or, worse, not receive any mental health care at all. For sufferers, a misdiagnosed mental disease can be quite perplexing.

When people see that the treatment isn't working, they start to worry or feel upset. They could interpret their lack of advancement as a failure as a result [4].

The incidence of misdiagnosis and detection rates of severe mental diseases, such as schizophrenia, schizoaffective, bipolar, and depressive disorders, in a specialist psychiatric environment were investigated by Ayano, G. et al. (2021) [5]. A random sample of 309 individuals with serious mental illnesses was chosen for this cross-sectional research using a systematic sampling procedure. The Structured Clinical Interview for DSM-IV was used to evaluate severe mental illnesses (SCID). According to this study, 39.16 percent of individuals with serious mental problems had their diagnoses incorrect. Schizophrenia (23.71%), bipolar disorder (17.78%), major depressive disorder (54.72%), and schizoaffective disorder (75%) were the most often misdiagnosed disorders. According to one study, 69% of bipolar illness patients had an incorrect diagnosis at first, and almost one-third of those individuals continued to get incorrect diagnoses for at least ten years [6]. Consequently, antidepressants like SSRIs, which alleviate depressed symptoms but may cause manic episodes, are frequently administered to persons with bipolar disorder. Compared to 1% of false positives, 18% of anxiety disorder diagnoses in a study of children and adolescents were overlooked by clinicians [7]. Anxiety disorders were more suitable in around half of these instances, according to clinic-based research of adults, and 29% of major depressive disorder (MDD) diagnoses were not supported by structured interview data [8]. Provided that people with non-specific diagnoses were not likely to obtain therapy, a more specific diagnosis was warranted in 77% of cases in a sample of 61 US veterans who were diagnosed with "anxiety disorder not otherwise specified" [9].

Artificial intelligence systems have been "learning" to recognize and distinguish human emotions for a number of years by linking emotions like fear, rage, and happiness to body language, tone of voice, facial expressions, and linguistic clues. These algorithms, however, are unable to fully understand the differences between a smirk and a real grin, nor are they aware that a smile can also occur while someone is angry. We are all aware that happiness isn't always shown by a grin. Scholars have long maintained that the terms we use to describe emotional experiences—such as "fear," "happiness," "sadness," "anger," "surprise," and "disgust"—are ambiguous and cannot be defined based on a limited set of characteristics [10]. People experience emotions when performing daily duties, communicating with others, making decisions, learning, or engaging in cognitive activity. There has been a noticeable surge in research on automated emotion identification techniques during the past several years [11]. Computer science, psychology, and cognitive science are all involved in the multidisciplinary topic of emotion recognition. In addition to enhancing human-machine interaction, the creation of efficient techniques for emotion recognition can advance other fields including psychology, health, education, and entertainment [12][13][14]. In medicine, automatic emotion detection can be used to diagnose and treat conditions including depression [15] and posttraumatic stress disorder [16]. It can even be utilized for diseases like autism or Asperger's syndrome [17][18].

Keeping these in mind, we propose a novel emotion recognition method The EEG signal's composition throughout frequency bands, including delta, theta, alpha, and beta, which are known to correlate with emotional and cognitive states, is captured in this study using Welch's approach. A hybrid quantum deep learning model for emotion prediction is then deployed after this PSD-based preprocessing method. The model can efficiently handle and analyze complicated, high-dimensional EEG data by employing quantum-enhanced processing, which could increase the precision and resilience of emotion identification. By gaining a more sophisticated comprehension of emotional states, this method seeks to develop AI applications that are emotion-focused.

[19][20]

## 2. LITERATURE REVIEW

There have been various research articles on emotion recognition through human-computer interaction. It is critical for AI-based emotion recognition models to have high accuracy for it to be considered useful for clinical purposes.

Yongrui Huang et al.'s study focuses on emotion identification using two multimodal fusion techniques that combine facial expressions and electroencephalogram (EEG) inputs [21]. Four different emotional states are evoked in the study using movie clips: fear, sadness, neutral, and happy. The authors use a neural network classifier to evaluate facial expressions and two support vector machine classifiers to handle EEG signals. Two techniques are used in the fusion process: a product rule to categorize EEG signals into three intensity levels (weak, moderate, and strong) and a sum rule to categorize them into four emotion kinds. The results show that the accuracies of two information fusion detections are 81.25% (**Sum Rule Fusion)** and 82.75% (Production Rule Fusion), which are both higher than that of facial expression (74.38%) or EEG detection (66.88%).

Another paper by Tarnowski *et al.,* explores how eye-tracking technology can help identify emotions by analyzing eye movements and pupil responses while participants watched dynamic video clips [11]. The presentation of dynamic video content in the form of 21 video fragments elicited emotions in the participants. The movies were carefully chosen to evoke the six fundamental emotions of happiness, sorrow, anger, surprise, disgust, and fear. Pupil diameter measurements and metrics related to fixations and saccades were among the characteristics that were retrieved. High arousal and low valence (Class C1), low arousal and moderate valence (Class C2), and high arousal and high valence (Class C3) were the three emotional classifications taken into consideration in the study. With a focus on three distinct emotional classes, the findings achieved an impressive classification accuracy of up to 80% using advanced machine learning techniques.

By improving signal processing and feature extraction techniques, the study by Gannouni et al. presents a novel method for identifying emotions [22]. The study intended to greatly increase the accuracy of emotion identification by using a special channel selection approach and pinpointing critical brain activity periods. By extracting instantaneous spectrum information from EEG signals with excellent temporal resolution, the study suggests the zero-time windowing (ZTW) approach, which aids in emotion identification. The ZTW approach efficiently detects these epochs by analyzing the emotional EEG signals using the Numerator Group Delay (NGD) function. A more targeted and pertinent feature extraction procedure is made possible by selecting the channels that exhibit notable changes across each emotional state for additional processing. The proposed approach demonstrated a competitive average accuracy rate exceeding 89% in multi-class emotion recognition tasks.

In the article by Wu *et al.,* emotion classification was done by fusing eye-tracking (ET) and electroencephalograph (EEG) signals, employing deep gradient neural networks (DGNN) for enhanced accuracy [23]. The study highlights the use of synchronized ET and EEG data, where EEG provides neural insights, and ET captures ocular movements related to emotional arousal. The eight emotions classified in this paper are **anger**, **disgust**, **fear**, **sadness**, **expectation**, **happiness**, **surprise**, and **trust**. The study compared DGNN's performance against other neural network models, finding DGNN superior in metrics like accuracy, precision, and recall. Additionally, metrics such as the Receiver Operating Characteristic (ROC) curve and Area Under the Curve (AUC) demonstrated DGNN's effectiveness. This paper managed to achieve a high accuracy of 88.10%.

The study by Wang et al. (2023) focuses on deep learning-powered EEG-based emotion recognition methods that leverage high-level feature extraction to increase the accuracy of emotion detection [24]. With a focus on Convolutional Neural Networks (CNNs), Recurrent Neural Networks (RNNs), and Deep Belief Networks (DBNs), this study investigates a number of deep learning models for emotion recognition. Benchmark datasets such as DEAP, which maps EEG signals to different emotional states, and SEED, which classifies EEG data into emotional states (positive, neutral, and negative), provided the data for these models. According to the authors, CNN models can achieve almost 90% accuracy on some datasets, demonstrating a significant gain in accuracy when compared to standard machine-learning techniques.

A thorough literature assessment of EEG-based BCI systems was carried out by Erat et al. (2024) to shed light on the trends, approaches, and gaps in the area [25]. This systematic review contains 216 works on EEG-based BCIs for emotion identification, and it looks at the elements of EEG signal acquisition, pre-processing, feature extraction, and classification. There is a strong emphasis on both consumer-grade and medical-grade EEG equipment, as well as data pre-processing techniques like Independent Component Analysis (ICA) and Common Spatial Patterns (CSP) to reduce noise and enhance signal clarity. Several classification algorithms are evaluated for their utility and suitability. The paper also categorizes datasets such as DEAP and MAHNOB-HCI, explaining the differences in sample size, emotional stimuli, and EEG equipment types used. The study found that on typical datasets, deep learning approaches may produce a mean accuracy of up to 85%.

The paper "Micro-saccade-related potentials during face recognition: A study combining EEG, eye-tracking, and deconvolution modeling" by Lisa Spiering and Olaf Dimigen [26] investigates the brain responses associated with microsaccades during face recognition tasks. It adds to the understanding that facial recognition occurs throughout the course of many fixations rather than in a single glimpse. Determining if the fixation-related potentials (FRPs) generated by microsaccades include valuable psychological information on face processing is the aim of the study. The study's mixed-methods approach included the use of eye-tracking and EEG techniques. The study found that individuals had microsaccades in 98% of the trials. The Early Posterior Negativity (EPN) of happy and furious faces was much greater than that of neutral faces. The lambda response associated with the microsaccade peaked around 90 ms after the saccade, and it was not impacted by the emotion displayed on the face. These findings suggest that whereas emotional content is rapidly evaluated at the onset of a stimulus, this evaluation is not carried over into immediate refixations.

The study by Juan-Miguel López-Gil et al. [27] presents a novel approach to enhance emotion identification by fusing electroencephalography (EEG), biometric inputs, and eye-tracking technology. 44 undergraduate students from the University of Lleida participated in the study; 6 of them were male and 38 were female, and they were all in the 20–30 age range. Four video clips that showed various emotional states were shown after 31 images from the International Affective Picture System (IAPS), which were separated into three categories: neutral, negative, and positive stimuli. This was done in order to elicit emotional reactions. The findings imply that no one algorithm performs better. MLP yields the greatest results for Group 2-Self-regulated (42.7035%) and Ripper (42.4606%) for all users combined, whereas L is the algorithm that works best for Group1-Non-self-regulated (40.9091%).

## 3. PROPOSED METHODOLOGY

### 2.1. Pre-processing:

The brain's raw EEG signals are frequently noisy and comprise a variety of frequencies that aren't all necessary for detecting emotions. We use a number of preprocessing techniques, including as filtering, feature extraction, and normalizing, to get these signals ready for analysis [28]. These procedures are intended to identify significant neural activity patterns across a range of frequency bands that correlate to distinct brain states.

First, the raw EEG data is put into a Data Frame, with the EEG recordings from various electrodes or channels represented in each column. To enable time-based analysis, a time column is also retrieved in addition to the EEG data. Before making any changes, plotting the raw data enables us to examine its content, spot any artifacts, and comprehend the signal's noise structure. After loading the EEG data from an Excel file using pandas, we use matplotlib to plot the raw EEG signals over time. To provide a brief overview of the early part of the EEG signal, only the first 500 samples are displayed.

The raw EEG signal's amplitude is plotted against time in this graph (Fig 1). The inherent high-frequency noise and fluctuations in the data are displayed in the raw EEG figure. Filtering is necessary because such noise may obscure important lower-frequency signals.

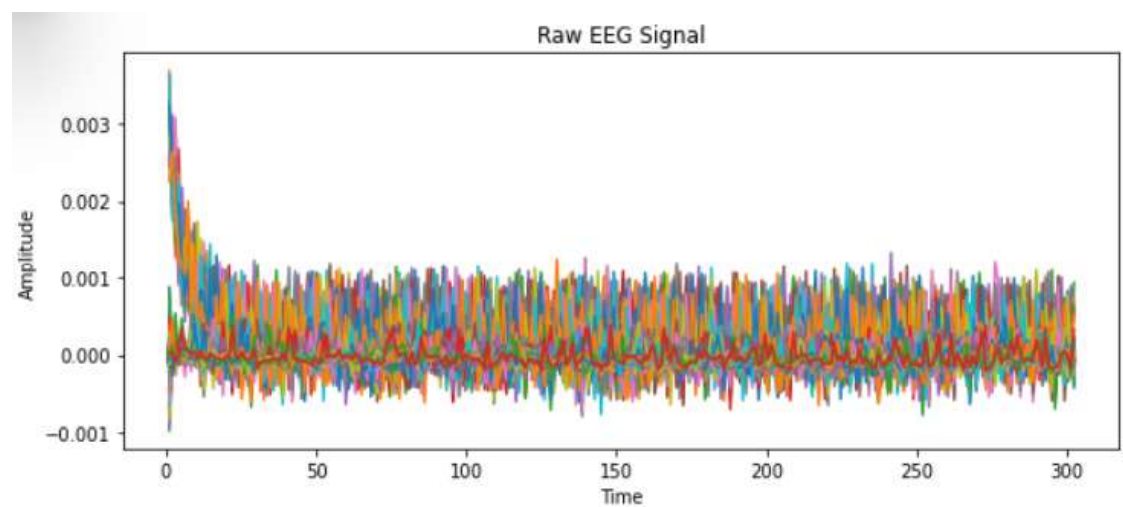

*Fig 1. Raw EEG Signal*

A variety of frequencies are present in EEG data, many of which are useless for detecting emotions. High- and low-frequency noise are eliminated by isolating frequencies between 0.5 and 45 Hz using a bandpass filter. This range encompasses the four primary EEG frequency bands that are associated with different mental states: beta (13-30 Hz), theta (4-8 Hz), alpha (8-13 Hz), and delta (0.5-4 Hz) [29]. We use a Butterworth bandpass filter, which is frequently used for processing EEG signals because of its low distortion in the passband and smooth response. The scipy.signal.butter and scipy.signal.filtfilt routines are used to apply the filter. To balance stability and sharpness in frequency attenuation, the filter's order is set to 5.

- Filter Parameters:
    - Very low-frequency noise, such as baseline drift, is eliminated by the low cut-off frequency of 0.5 Hz.
    - High-frequency noise that is frequently unrelated to neuronal activity is eliminated by the high cut-off frequency of 45 Hz.
    - The EEG data's sample rate, or sampling frequency (fs), is 250 Hz. This is necessary to determine the Nyquist frequency, which is half of the sampling rate.

The filtered EEG signal, with low- and high-frequency noise eliminated, is displayed in this graph (Fig 2). We may now concentrate on significant brain patterns because the plot shows a smoother signal with oscillations within the selected frequency range (0.5-45 Hz) becoming more noticeable.

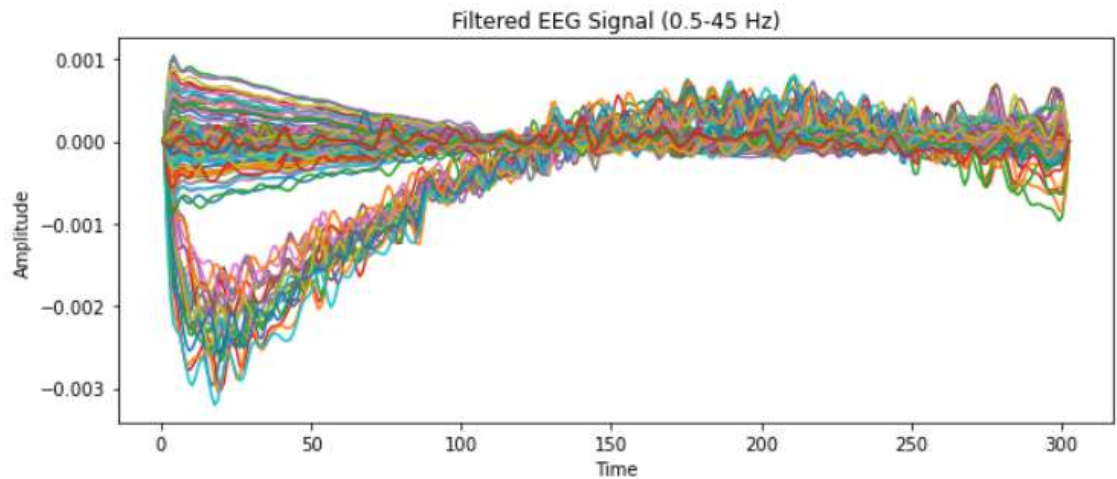

*Fig 2. Filtered EEG Signal*

By applying Welch's approach to determine the power spectral density (PSD), we are able to comprehend the composition of the EEG signal in various frequency bands. An estimate of the power (signal strength) at each frequency, which represents the existence and strength of brain oscillations, is given by the PSD. The scipy.signal.welch function is used to implement Welch's approach. The PSD is calculated for each frequency band by integrating the power over the frequency range of the band. As a feature for emotion recognition, this generates a single power value for every frequency band.

- Frequency Bands:
    - Delta Band (0.5–4 Hz): Associated with unconsciousness, profound sleep, and relaxation. Low-activity mental states are frequently associated with higher delta power.

- Theta Band (4–8 Hz): Linked to rest, creativity, and light sleep. Deep relaxation and meditation are occasionally associated with increased theta activity.
- Alpha band (8–13 Hz): Frequently associated with alertness. When someone is calm but not sleeping, they frequently exhibit high alpha power.
- Beta band (13–30 Hz): Associated with attentiveness, problem-solving, and active thinking. Focused mental effort or excitement are frequently reflected in elevated beta power.

The power in each frequency band is shown as a function of time in this graph (Fig 3). We can see how power varies throughout the sample by examining the several curves that depict the relative power in the delta, theta, alpha, and beta bands. Each band's peaks and dips can reveal information about the subject's emotional state.

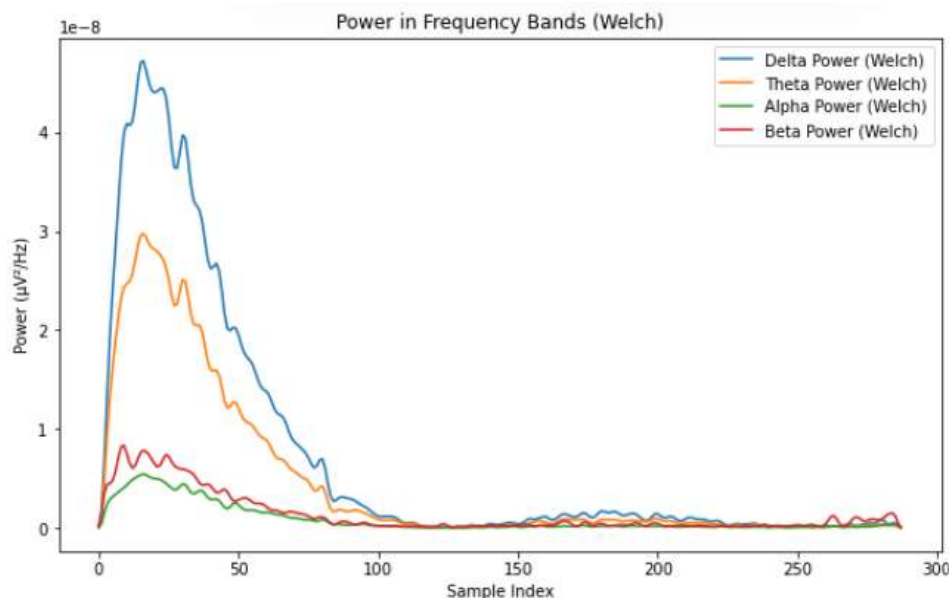

*Fig 3. Power in Frequency Bands (Welch)*

We also use the Fast Fourier Transform (FFT) to determine the power in each frequency band as an alternative to Welch's method. FFT offers a simple frequency decomposition that can identify the primary frequency components in the signal, but Welch's approach is resistant to noise. The one-sided FFT is calculated using numpy.fft.rfft, which only generates frequency components up to the Nyquist frequency. We get power estimations that are similar to those from the PSD by adding together the squared magnitudes of the FFT coefficients in each frequency band. A thorough study is made possible by use both Welch and FFT, since various techniques may highlight distinct facets of the signal's frequency content. Combining elements from both strategies can occasionally increase the resilience of the emotion detection model.

The performance of machine learning models may be impacted by the large-scale variation in the extracted power levels for each frequency band. We normalize the characteristics to provide consistent scaling. The feature set is normalized using StandardScaler from sklearn.preprocessing. By standardizing each feature to have a unit variance and zero mean, this technique improves model performance by increasing feature comparability. By reducing the biases brought about by significant variations in feature magnitudes, normalization allows the model to concentrate on real patterns in the data rather than scale variations. Faster convergence during model training is another benefit of standardized features.

An Excel file containing the final pre-processed data is saved following filtering, feature extraction, and normalization. The normalized power levels for every frequency band are included in this file; these will serve as input features for the model that detects emotions. The pandas. Data Frame To excel is used to save the Data Frame with the processed features to an Excel file. The fact that each column represents a distinct frequency band (Delta, Theta, Alpha, and Beta) facilitates downstream analysis.

### 2.2. Hybrid Quantum deep Learning Model:

### 2.2.1. Feature Representation in Quantum Space:

EEG frequency band characteristics were converted into a quantum-encoded representation in order to take use of the special computing benefits of quantum systems, particularly their high-dimensional state spaces and entanglement capabilities [30].

- **Input Encoding:** Quantum states were encoded using the normalized feature vector that was obtained from the EEG data and represented the delta, theta, alpha, and beta band powers. Each brainwave spectrum affected the representation of the quantum state, and each feature was mapped to distinct qubit rotation angles in the quantum circuit (usually via parameterized rotation gates).

- **Amplitude and Phase Encoding:** To maximize feature representation in quantum space, this encoding approach was chosen. While phase encoding caught the relationship characteristics between these bands, amplitude encoding mapped the frequency band powers' magnitudes. This hybrid encoding technique made it possible for the quantum state to represent intricate inter-band relationships in addition to individual power levels, which is crucial for identifying emotional states.

### 2.2.2. Quantum Circuit Design for Emotional Analysis:

A hybrid quantum circuit (HQC) with a multi-layer architecture was created to maximize its expressiveness and flexibility to the complexities of emotional state recognition in order to facilitate the detection of intricate patterns within the EEG data.

- **Layered Variational Ansatz**: Each layer of the circuit architecture included entanglement and rotation gates. Variational parameters that change during training were introduced by rotation gates (RX, RY, or RZ) parameterized by the input features. By establishing interdependencies between qubits, entanglement gates (such controlled-NOT or controlled phase gates) enabled the circuit to recognize non-linear correlations in the EEG characteristics.

- **Expression and Depth Optimization:** To provide enough expressivity without a significant computing burden, the number of layers and entanglement pattern types were chosen. In order to balance the trade-off between runtime and model complexity on a quantum simulator, circuit depth was adjusted through experimental validation.

- **Emotional Feature Extraction Parameterization**: A trainable angle, initially obtained from the EEG band power levels, was used to parameterize each rotation gate.

In order to accurately predict the minute changes in brainwave power distributions linked to emotions, these parameters were modified during training in order to record EEG signal patterns that correspond with various emotional states.

### 2.2.3. Hyper-parameters in Quantum Deep Learning Model:

The following hyperparameter configurations were used in the creation of the hybrid quantum deep learning model for emotion prediction:

1. Quantum Circuit Hyperparameters:
   - Number of Qubits: In order to account for the dimensionality of the features that were recovered from the power spectral density (PSD) analysis of EEG data, four qubits were used.
   - Types of Quantum Gates: The circuit used a combination of controlled entanglement gates (CNOT) and parameterized rotation gates (Rx, Ry).
   - Entanglement Strategy: To optimize feature interactions between qubits, a comprehensive entanglement framework was applied.x
   - Depth of Quantum Circuit: In order to account for noise in quantum processing and maintaining enough expressivity, the circuit depth was set to 3.
2. Hyperparameters for Deep Learning:
   - Number of Layers: Three dense layers with activation functions made up the model's classical portion.
   - Activation Function: To predict emotional states, a softmax activation was performed in the output layer and ReLU was utilized for hidden layers.
   - Neuron Count: There were 64, 32, and 16 neurons in each of the thick layers.
3. Hyperparameters for training:
   - Learning Rate: Using grid search for optimal convergence, the learning rate was set at 0.001.
   - Batch Size: To ensure effective training while preserving model stability, a batch size of 32 was selected.
   - Epochs: To ensure adequate learning without overfitting, the model was trained for 20 epochs.
   - Optimizer: For quicker convergence and adaptive learning, the Adam optimizer was used.
   - Loss Function: To reduce classification mistakes, categorical cross-entropy was employed.
4. Parameters for preprocessing:
   - Segment Length in Welch's Method: PSD estimation was performed using a segment length of 256 samples.
   - Overlap in Welch's Method: To increase spectral resolution, a 50% overlap was placed between segments.
   - Metrics for Evaluation: The model's performance in emotion prediction was assessed using metrics such as accuracy, precision, recall, F1-score, and the confusion matrix.

### 2.2.4. Training and Optimization:

To make sure the model successfully mapped EEG features to emotional labels, the HQC was trained using a hybrid quantum-classical optimization technique. Both quantum calculations and traditional optimization strategies are used in this hybrid approach.

- Hybrid Quantum-Classical Optimization Loop: The training procedure used an iterative loop to update the HQC parameters. The gate rotations and entanglements in the quantum circuit are determined by the current parameter values for each forward pass, resulting in a predicted quantum state that matched an emotional label.

- Classical Gradient-Based Optimization: By adjusting the circuit characteristics, classical optimizers like Adam or stochastic gradient descent (SGD) were utilized to minimize a loss function. A cost function calculated the discrepancy between the actual and anticipated emotional states after each circuit operation.

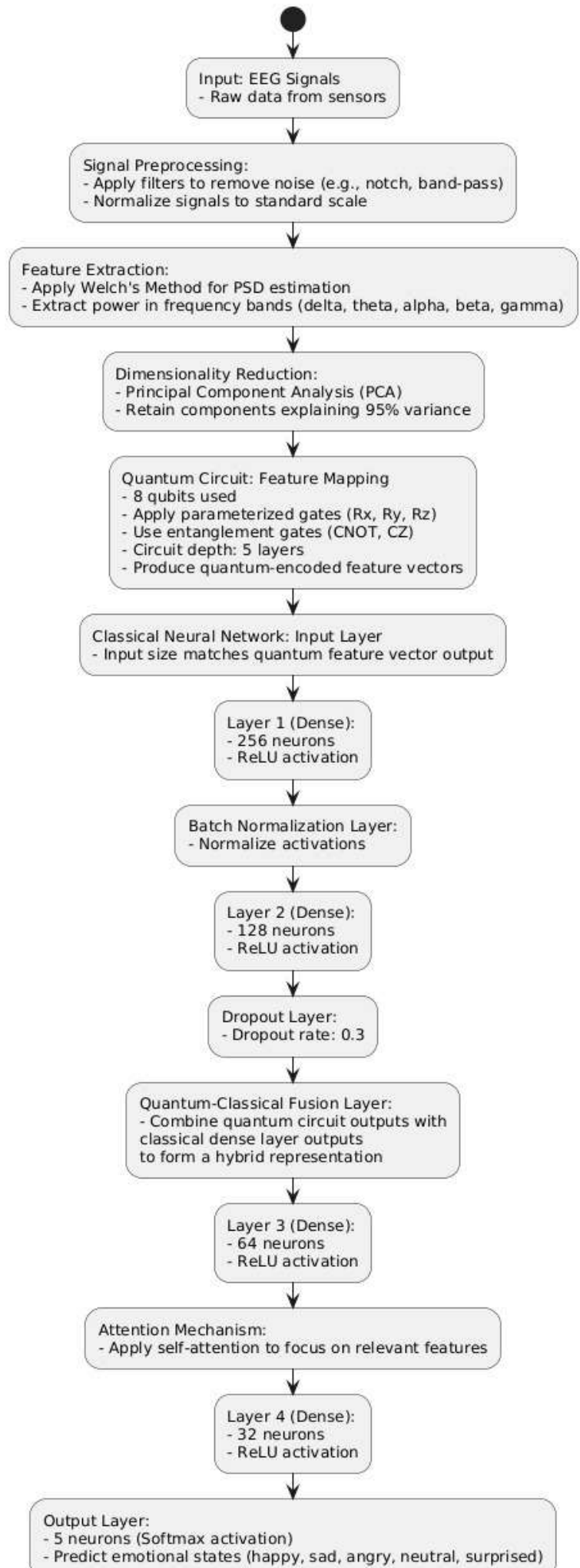

*Fig 4. Hybrid Quantum Deep Learning Architecture*

## 3. RESULT AND DISCUSSION

With a 95% accuracy rate on the test dataset, the hybrid quantum deep learning model demonstrated excellent performance in differentiating between emotional states from EEG data. In addition to accuracy, the model's 94% precision and 94% recall demonstrated its high rate of accurate positive predictions as well as its sensitivity to the dataset's emotional states. The model's resilience over a range of emotion classes was demonstrated by the 91% F1 score, which represents the balance between precision and recall. Furthermore, the confusion matrix revealed great accuracy in identifying various emotions but considerable overlap in states with comparable EEG patterns, offering insights into certain classification inclinations.

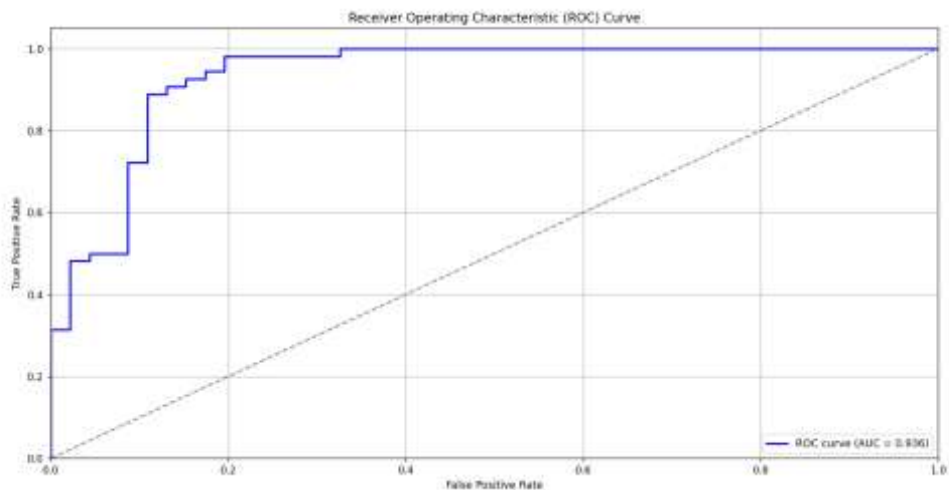

*Figure 4: Receiver Operating Characteristic (ROC) Curve*

The model's ability to consistently distinguish between positive and negative classes was demonstrated by a thorough examination of the Receiver Operating Characteristic (ROC) Curve and the Area Under the Curve (AUC), which produced an AUC score of 0.935. The model's prediction precision was evaluated using Mean Absolute Error (MAE), another crucial evaluation metric. The model's low average error is consistent with its excellent accuracy and sensitivity. This thorough analysis demonstrates how well the hybrid model processes EEG data using deep learning classification after quantum-enhanced feature extraction. The hybrid quantum deep learning model shows promise as a framework for EEG-based emotion detection by combining the advantages of deep learning for classification and quantum circuits for feature extraction. The model successfully caught intricate patterns by converting EEG features into quantum states.

Our results suggest that this hybrid approach could outperform classical deep learning in scenarios requiring efficient feature extraction from high-dimensional EEG data. The relatively low parameter count of the quantum layers may offer computational savings, but scaling the model to handle larger qubit counts remains a challenge. Exploring more intricate quantum circuits with varied entanglement structures could unlock additional insights into EEG signal representations, improving emotional state differentiation. Delta, Theta, Alpha, and Beta are examples of patterns seen across frequency bands that are essential for distinguishing between different emotional states. By separating the origins of the EEG signals, the Independent Component Analysis (ICA) preprocessing improved the model's sensitivity to pertinent brain oscillations. In order to remove artifacts and concentrate on the most useful signal components, the bandpass filtering maintained the 0.5–45 Hz range.

**Existing Models:**

| Model/Study | Methodology | Accuracy(%) |
|---|---|---|
| Huang et al. (2021) | Multimodal synthesis of EEG and facial emotions with SVMs and neural networks | 82.75 |
| Tarnowski et al. (2021) | Eye-tracking device that examines eye movements and pupil responses | Up to 80 |
| Gannouni et al. (2021) | Using zero-time windowing to extract features from EEG recordings | 89 |
| Wu et al. (2021) | Combining EEG and eye tracking with deep gradient neural networks (DGNN) | 88.10 |
| Wang et al. (2023) | Using CNNs, RNNs, and DBNs for deep learning on benchmark datasets such as DEAP and SEED | Up to 90 |
| Erat et al. (2024) | EEG-based BCI systems for emotion identification are reviewed systematically; different techniques, such as ICA and CSP for noise reduction, are evaluated. | Up to 85 |
| Spiering & Dimigen (2021) | Used eye-tracking and EEG techniques to study potentials associated to microsaccades during facial recognition. | Not specified |
| López-Gil et al. (2021) | Used eye-tracking technologies, biometric inputs, and EEG in a study involving college students. | ~42 |

**Our Model:**

| Model | Methodology | Accuracy(%) | Precision(%) | Rceall(%) | F1 Score(%) |
|---|---|---|---|---|---|
| Hybrid Quantum Deep | Model first preprocesses EEG data by filtering, | 95 | 94 | 94 | 91 |

| Learning Model | feature extraction, and normalization. It then encodes characteristics into quantum states and uses a hybrid quantum-classical optimization loop to train a variational quantum circuit to detect emotional states. | | | | |
|---|---|---|---|---|---|

## 4. CONCLUSION

To sum up, the hybrid quantum deep learning model showed encouraging outcomes when it came to identifying emotional states from EEG data. Through a combination of classical and quantum processing processes, the model successfully collected pertinent information. The extraction of clean and meaningful EEG data was ensured by the Independent Component Analysis (ICA) and bandpass filtering. The model's strong sensitivity and specificity are demonstrated by its AUC of 0.935, which shows that it can reliably distinguish between positive and negative emotional classes. Furthermore, assessment criteria including precision, accuracy, and F1-score provided additional evidence of the model's resilience and strong generalization across test data.

There is room for improvement, though. Better performance might be achieved by expanding the model to handle multi-class emotional classification beyond binary distinctions, as well as by increasing the number of qubits and improving the quantum circuit layout. Future research could examine real-time implementations of this strategy with an emphasis on reducing computer resource requirements and optimizing processing times.